\documentclass[a4paper,12pt,twocolumn]{article} 
\usepackage{graphicx}
\usepackage{multirow}

\newcommand{\beq}{\begin{eqnarray}}
\newcommand{\eeq}{\end{eqnarray}}

\begin{document}
\pagestyle{empty}

\title{
{\Large \bf 
Single shot emittance measurement using Optical Transition Radiation
}
}
 
\author{
{\small 
Nicolas Delerue, Riccardo Bartolini and Cyrille Thomas
}
}

\date{
November 2008
}

\maketitle
\thispagestyle{empty}

\begin{center}
{\bf 





We describe a method that uses Optical Transition Radiation (OTR) screens to measure in a single shot the emittance of an electron beam with an energy greater than 100 MeV. Our setup consists of 4 OTR screens located near a beam waist. A fit of the 4 profiles allows the reconstruction of the Twiss parameters and hence a calculation of the beam emittance. This layout has been simulated using Mathematica to study its sensitivity to errors and using Geant4 to include effects due to the scattering of the electrons in the screens.\\
Some assumptions made in this document are conservative to allow for unexpected experimental issues.

}
\end{center}

\section{Emittance measurement using several screens} 

\subsection{Proposed experiments}

The measurement of a beam emittance by using several screens one after the other in a linac is a well know technique. We propose to adapt this technique to single shot emittance measurements by using the screens simultaneously. This requires the use of very thin screens. In our simulations we have used 3 screens located 0.5m apart from each other followed by a 4th screen located 1m beyond the 3rd screen.

\begin{figure*}[htbp]
\begin{center}
\includegraphics[height=4cm]{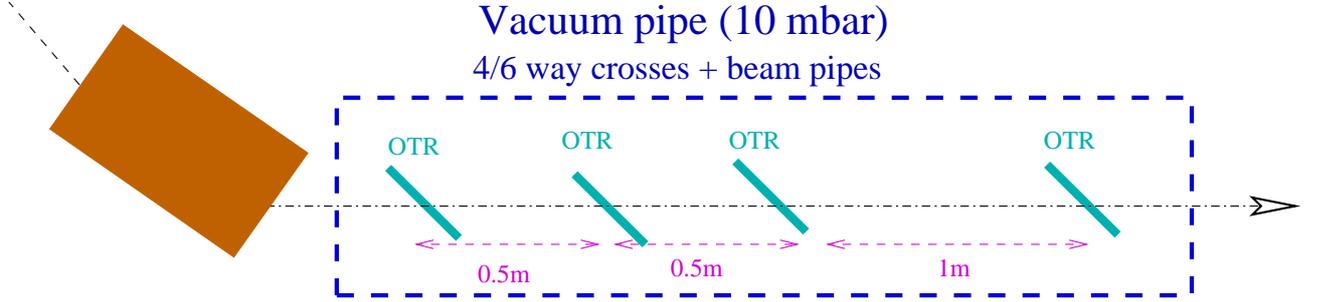} 
\caption{
Proposed layout
\label{fig:layout} }
\end{center}
\end{figure*} 

\subsection{Beam propagation in a drift space}

In a drift space the beam ($\sigma$) size evolves as~\cite{CAS94}:

\begin{equation}
\sigma(s) = \sqrt{\epsilon (\beta_0 - 2 \alpha_0 s + \gamma_0 s^2)}
\label{eq:beamDriftGnrl}
\end{equation}

where $s$ is the position in the drift space, $\alpha_0$, $\beta_0$ and $\gamma_0$ are the Twiss parameters at $s=0$ and $\epsilon$ is the beam emittance.

When $s=0$ is a symmetry point (waist), this can be rewritten as

\begin{equation}
\sigma(s) = \sqrt{\epsilon \beta_0 + \frac{\epsilon}{\beta_0}s^2}
\label{eq:beamDrift}
\end{equation}

By measuring $\sigma$ with an OTR screen at 3 (or more) locations around a waist it is thus possible to make a parabolic fit giving the Twiss parameters of the beam as well as 
$\epsilon$ and $\beta_0$.

\subsection{Beam expansion condition}

In such drift space the beam divergence is (with a different sign on either side of the waist)

\beq
\sigma' = \pm \sqrt{\frac{\epsilon}{\beta}} =  \pm \frac{\epsilon}{\sigma_0} \label{eq:divergence}
\eeq

with $\sigma_0$ the beam size at the waist.

The beam size at a given distance $s$ from the waist will be related to the beam size at the waist by the following relation:
\beq
\frac{\sigma(s)}{\sigma_0} & = & \frac{\sqrt{\epsilon \beta_0 + \frac{\epsilon}{\beta_0} s^2}}{\sqrt{\epsilon \beta_0}} \nonumber \\
& = & \sqrt{1 + \frac{s^2}{\beta_0^2}}
\eeq

Using $ \beta_0 = \frac{\sigma_0^2}{\epsilon} $, this becomes:

\beq
\frac{\sigma(s)}{\sigma_0} & = & \sqrt{1 + \frac{\epsilon^2 s^2}{\sigma_0^4}}
\label{eq:expansionCondition}
\eeq

\begin{figure*}[htbp]
\begin{center}
\includegraphics[height=5cm]{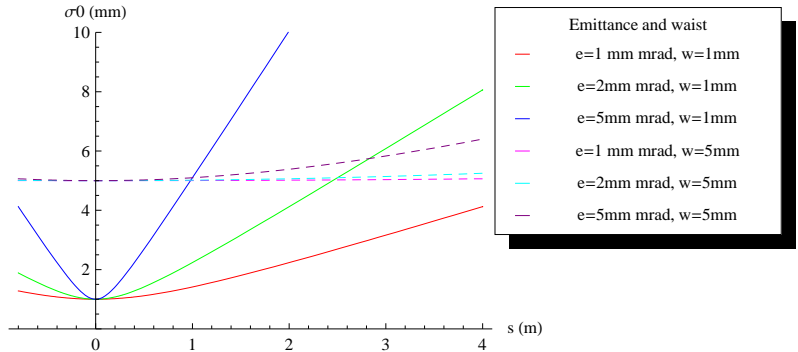} 
\caption{Beam envelope for different emittance and waist size values. Plain lines are for beams forming a 1~mm waist and dashed lines are for beam forming a 5~mm waist.
\label{fig:beamProfile} }
\end{center}
\end{figure*}

Hence for a 1~mm.mrad beam with a 1~mm spot size, the beam will have increased by 10\% beam at $s=0.45$ m and by 20\% at $s=0.66$m. For the same beam with a 5mm spot size, these distances are multiplied by 25 and become $s=11.46$m and $s=16.6$m. Figure~\ref{fig:beamProfile} gives the beam enveloppes for various emittances and beam waist sizes.

\subsection{Effect of the scattering in the screens on the beam expansion}

An electron traversing nuclear matter will be Coulomb scattered. Most of these scattering will be at very small angle but will add up as the electron progresses through the matter. The phenomena is described in details in~\cite{PDG}.

For a bunch of electron, the scattering experienced after traversing a layer of thickness $x$ can be approximated by a Gaussian of width\cite{Lynch:1990sq}:

\begin{equation}
\theta_0 = \frac{13.6\mbox{ MeV}}{\beta c p } \sqrt{\frac{x}{X_0}} \left[ 1+ 0.038 ln\left(\frac{x}{X_0}  \right)\right]
\label{eq:theta0}
\end{equation}

 where $\beta c$ is the velocity of the electrons, p their momentum and $X_0$ a quantity called the radiation length and which expresses how high energy electrons loose their energy in a given material\footnote{The radiation length is the mean distance over which a high-energy electron looses all but 1/$e$ of its energy by Bremstrahlung\cite{PDG}.}.

To make it easier to estimate the scattering as a function of the momentum this equation can be rewritten as 

\beq
p \theta_0 = \frac{13.6\mbox{ MeV}}{\beta c } \sqrt{\frac{x}{X_0}} \left[ 1+ 0.038 ln\left(\frac{x}{X_0}  \right)\right]
\label{eq:ptheta0}
\eeq

Table~\ref{table:nuclearProperties} shows that the electrons will travel only a small fraction of a radiation length ($X_0$) in each type of screen considered. A scattering factor $p \theta_0$ of the order of a hundred MeV.mrad  means that a beam with an energy of a hundreds MeV will be scattered by about one milliradian.
Among the different screens considered, Mylar and Kapton (Polyimide film) introduce the lower scattering and thinner films seems available.

\begin{table*}[htbp]
\begin{tabular}{l||c||c|c||c||}
&
Radiation  
& Thickness
& Scattering 
& Limit $\sigma_0$ (eq.~\ref{eq:cond_theta}) 
\\
\cline{3-4} 
Material & 
Length ($X_0$)
& $10^{-4}~X_0$  &  $p \theta_0$ 
& $\frac{1 \mbox{mm.mrad} }{\theta_0} $
\\
\hline
\hline

Aluminium (10 $\mu$m) &
8.9 cm &
1.12 & 
139~MeV.mrad &
0.9 mm
\\

Aluminium (30 $\mu$m) &
8.9 cm &
3.37 & 
242~MeV.mrad &
0.5 mm
\\

Mylar (2 $\mu$m) &
28.6 cm &
0.069 (*) & 
34~MeV.mrad &
3.7 mm
\\

Polyimide film (7.5 $\mu$m) &
28.6 cm &
0.26 (*) & 
66~MeV.mrad &
1.9~mm
\\

Polyimide film (10 $\mu$m) &
28.6 cm &
0.34 (*) & 
77~MeV.mrad &
1.6~mm
\\

Mylar (10 $\mu$m) &
28.6 cm &
0.34 (*) & 
77~MeV.mrad &
1.6~mm
\\

\hline
Air (2 m), 1~atm. &
3 $10^4$ cm &
66.67 & 
1089~MeV.mrad &
 - 
\\

Air (2 m), 0.01~atm. &
3 $10^6$ cm &
0.67 (*) & 
107~MeV.mrad &
 -
\\

\hline
\hline

\end{tabular}
\caption{Nuclear properties of various materials~\cite{PDG} for different thicknesses. (*) Equation~\ref{eq:ptheta0} is not fully valid at these thicknesses.}
\label{table:nuclearProperties}
\end{table*}

When the beam scatters on the screens an additional scattering term is required and equation~\ref{eq:beamDriftGnrl} is approximately modified as follow:

\begin{equation}
\sigma(s) = \sqrt{\epsilon (\beta_0 - 2 \alpha_0 s + \gamma_0 s^2)} + \sum^{N(s)_{\mbox{screens}}}_{i=1} (s-s_i) \theta_i^2
\label{eq:beamDriftScattering}
\end{equation}

where $N(s)_{\mbox{screens}}$ is the number of screens located upstream from the position at which the measurement is performed, $s_i$ their position and $\theta_i$ the scattering they induce.

Assuming that all screens are made of the same material with a scattering coefficient $\theta_0$, the scattering induced by the first screen will dominate:

\begin{equation}
\sum^{N(s)_{\mbox{screens}}}_{i=1} (s-s_i) \theta_i \simeq (s - s_1) \theta_0 
\end{equation}

A measurement of the beam size dominated by the term $(s - s_1) \theta_0 $ will provide information on the scattering but not on the emittance of the beam. Hence the screens, their position and the beam optics must be chosen so that in all measurements the scattering can be neglected.

This requires:

\beq
\frac{\epsilon}{\sigma_0} s > N_{\mbox{screens}} \theta_0 (s - s_1)
\label{eq:cond_theta_s}
\eeq

where $N_{\mbox{screens}}$ is the total number of screens used.

Assuming that the first screen is close from the waist and hence $s-s_1 \simeq s$, this can be rewritten as:

\beq
\frac{\epsilon}{\sigma_0} >> N_{\mbox{screens}} \theta_0 \\
\sigma_0 << \frac{\epsilon}{N_{\mbox{screens}} \theta_0} \label{eq:cond_theta} 
\eeq


Equation~\ref{eq:cond_theta} can be rewritten as

\beq
\sigma_0 << N_{\mbox{screens}} \frac{\epsilon_n}{\gamma \frac{  p \theta_0}{p}} \\
\sigma_0 << N_{\mbox{screens}} \frac{m c \epsilon_n}{ p \theta_0} 
\eeq

where $\epsilon_n$ is the normalised emittance, $\gamma$ the Lorentz transformation factor and c the speed of light.

This shows that the requirement on $\sigma_0$ as a function of the normalised emittance and of $p \theta_i$ is independent of the beam energy (but the ease of achieving a given $\sigma_0$ may be different). 

At 500~MeV a $7.5 \mu$m polyimide film screen will induce a scattering $\theta_i = 0.13 $mrad. With an emittance  $\epsilon = 1 $mm.mrad, equation~\ref{eq:cond_theta}  requires $\sigma_0 << 7.6 $mm. 
The limit value of $\sigma_0$ as given by equation~\ref{eq:cond_theta} assuming $N_{\mbox{screens}}=4$ for various screen materials is given in table~\ref{table:nuclearProperties}. 

When the spot size is chosen so that equation~\ref{eq:cond_theta} is satisfied the scattering can be neglected, it is the case with thin Mylar or kapton screens .

\section{Simulations} 
\subsection{Mathematica} 

Simple simulations of such measurement have been done using Mathematica. In these simulations 4 screens were positioned at $s=-0.5m$, $s=0m$, $s=0.5m$ and $s=1.5m$. The propagation of a beam with a given emittance and beta function was then calculated and the size of the beam at each screen was estimated with some noise added. To take into account the limited pixel size the measured beam size was rounded to the size of the pixels.

If the beam optics or the beam energy is not well understood the beam waist may not be positioned exactly where it is expected to be, this was included in our simulations.

Equation~\ref{eq:beamDrift} was fitted to the beam sizes found and from this fit the value of the emittance was deducted. With the beam sizes used it is assumed that the effect of the scattering can be neglected as the natural beam divergence is much bigger than the divergence induced (equation~\ref{eq:cond_theta}).

\begin{figure*}[htbp]
\begin{center}
\hspace*{-2cm}\begin{tabular}{cc}
(a) $\epsilon = 1$mm.mrad
&
(b) $\epsilon = 2.5$mm.mrad
\\
$s = 0 $m
&
$s = 0 $m
\\
\includegraphics[height=4cm]{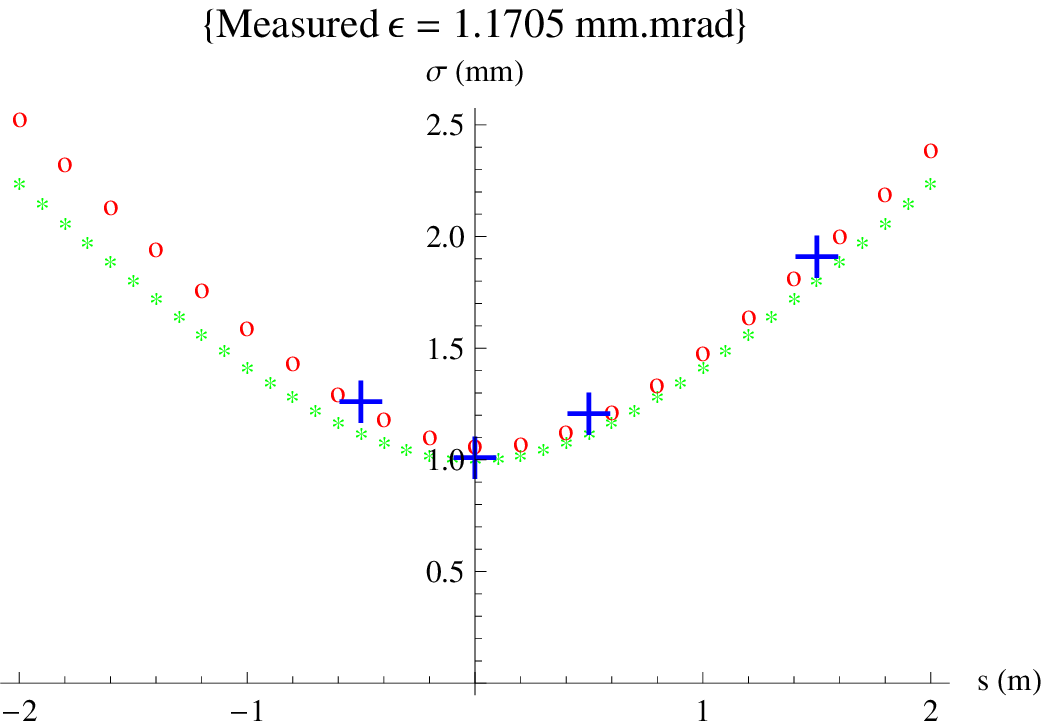} 
&
\includegraphics[height=4cm]{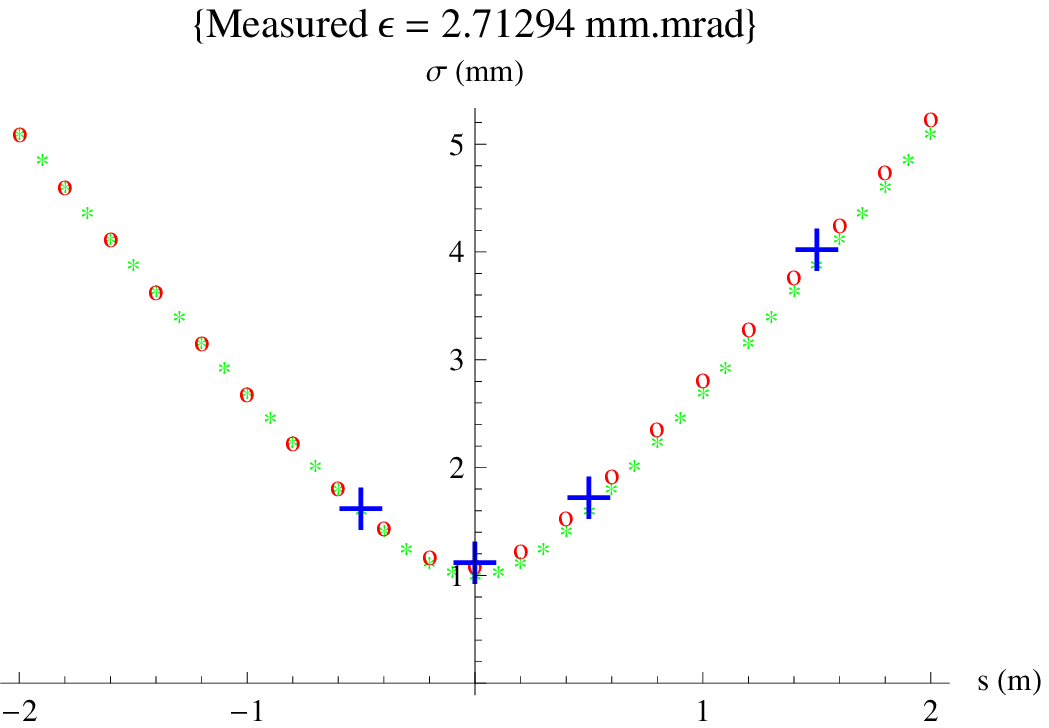} 
\\
(c) $\epsilon = 1$mm.mrad
&
(d) $\epsilon = 5.1$mm.mrad
\\
$s = 0.5 $m
&
$s = 0 $m \\
\includegraphics[height=4cm]{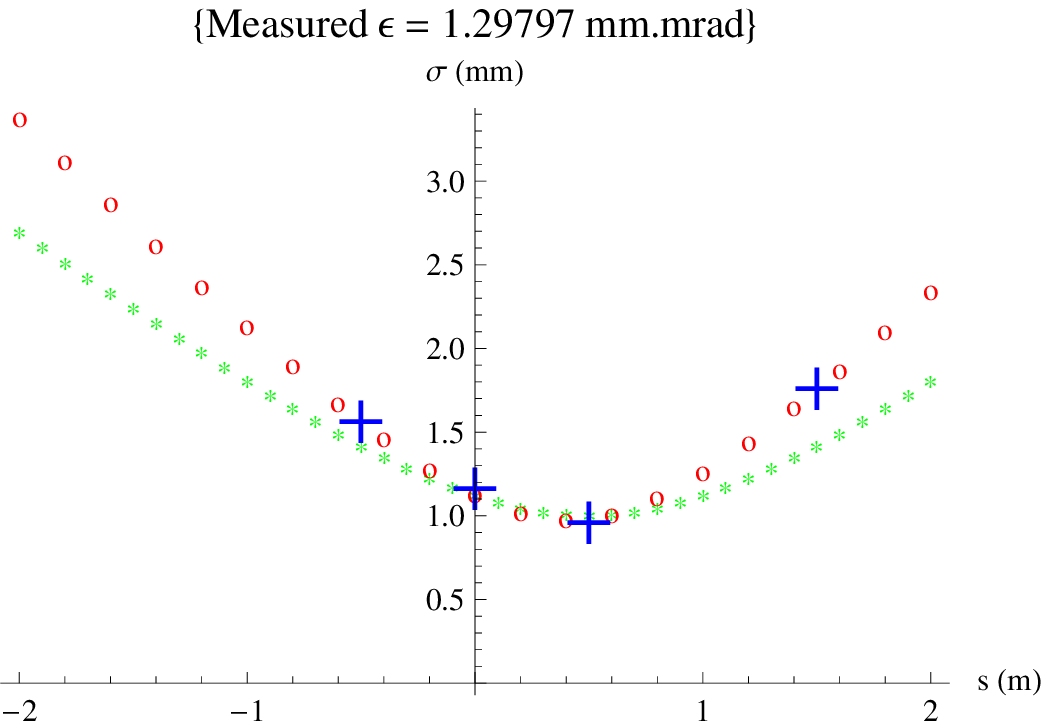} 
&
\includegraphics[height=4cm]{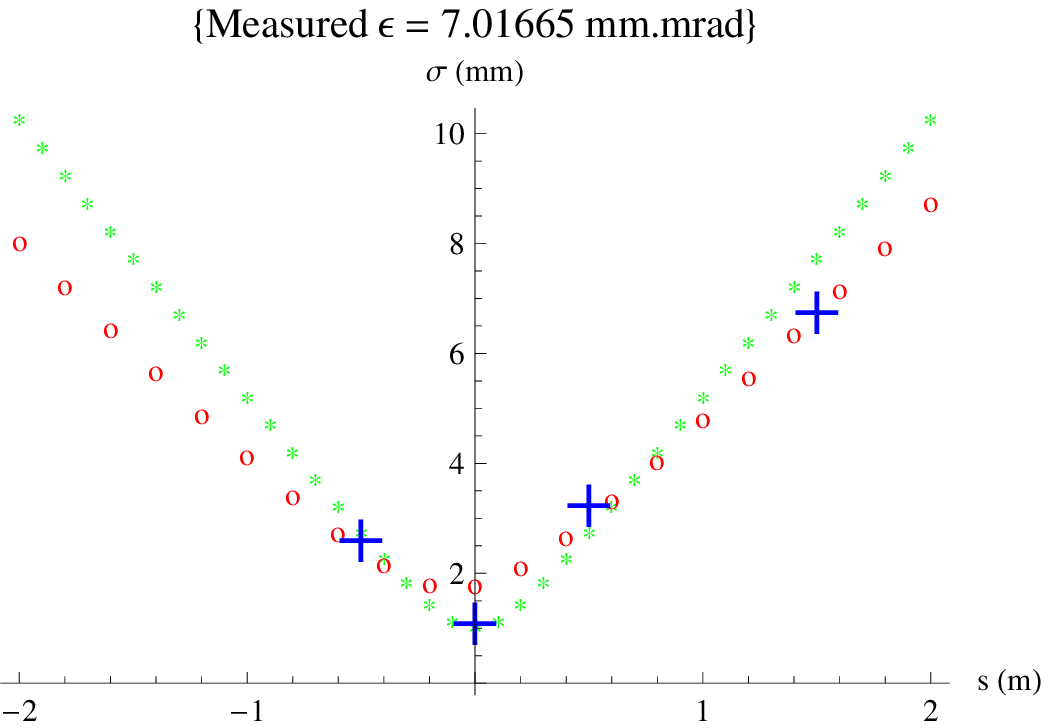} 
\\
\end{tabular}
\caption{Simulated (small green dots), measured (blue crosses) and fitted (red circles) beam sizes ($\sigma$) around the waist. The beam transverse emittance ($\epsilon$) and waist offset ($s$) is indicated above each plot. The waist size is always assumed to be 1~mm and the pixel size $20 \mu m$. The screens used are assumed to satisfy equation~\ref{eq:cond_theta} and thus the effect of the scattering is neglected in these plots.
\label{fig:emitFits} }
\end{center}
\end{figure*}

Figure~\ref{fig:emitFits} shows a few example of such fits.

\begin{figure*}[htbp]
\begin{center}
\hspace*{-2cm}\begin{tabular}{cc}
(a) Pixels = 50 $\mu$m & (b) Pixels = 50 $\mu$m \\
Noise = 5\% & Noise = 10\% \\
\includegraphics[height=4cm]{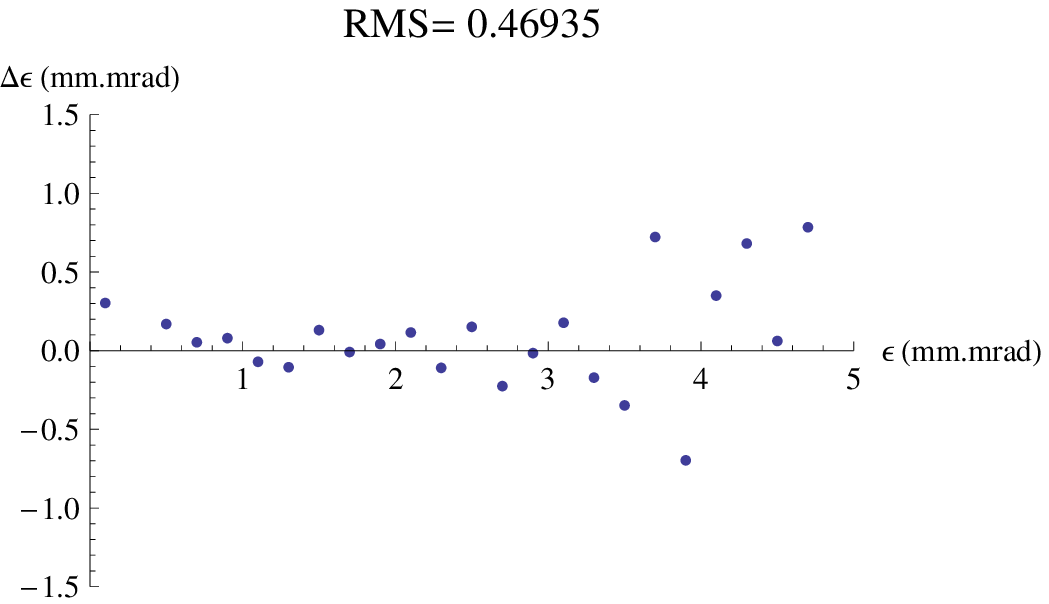} 
&
\includegraphics[height=4cm]{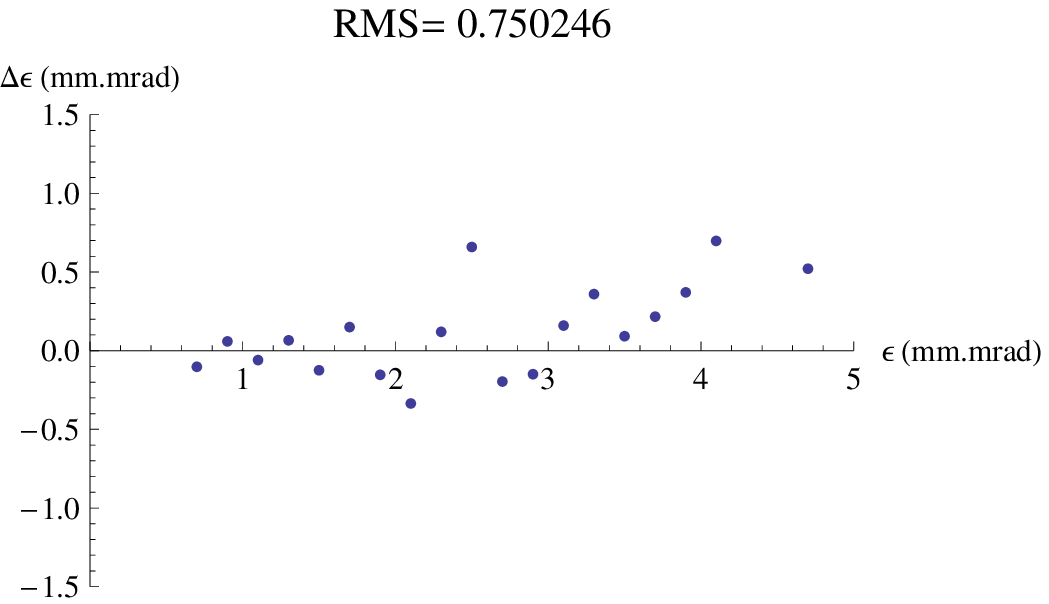} 
\\

(c) Pixels = 100 $\mu$m & (d) Pixels = 100 $\mu$m \\
Noise = 5\% & Noise = 10\% \\
\includegraphics[height=4cm]{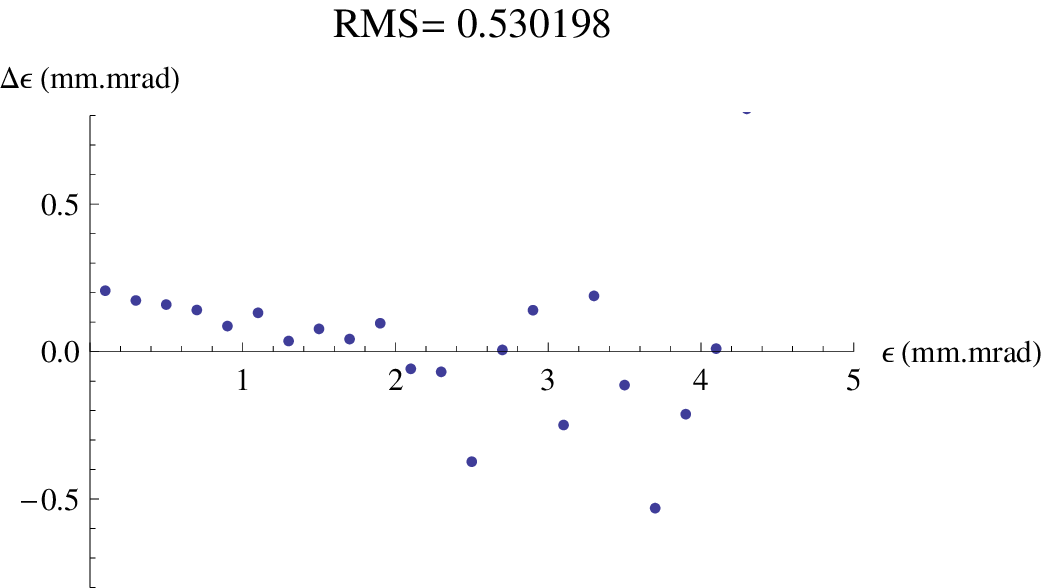} 
&
\includegraphics[height=4cm]{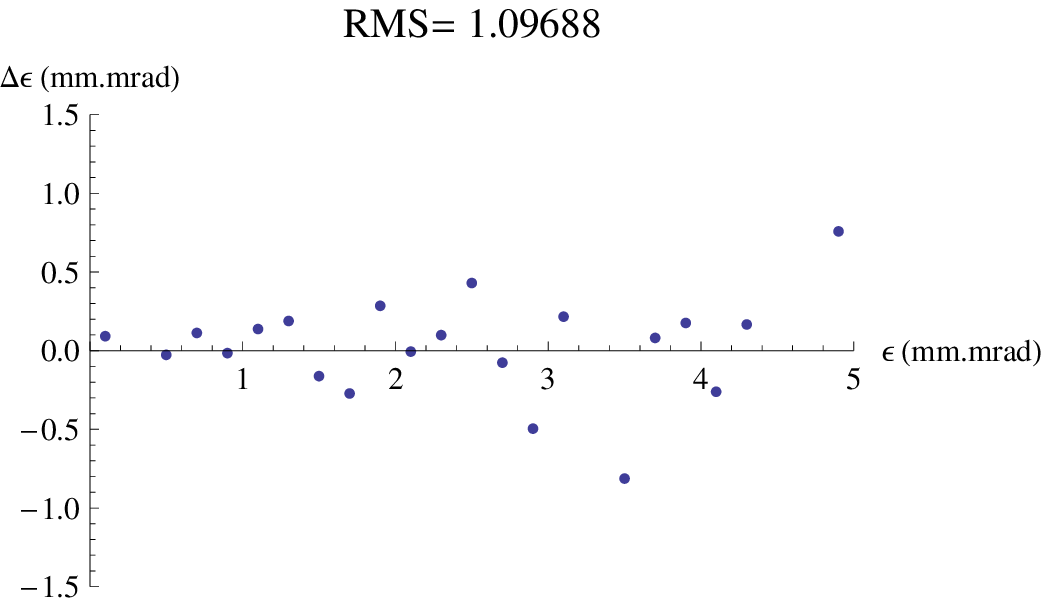} 
\\
\end{tabular}
\caption{Error on the emittance measurement as a function of the beam emittance for beams forming a 1~mm waist at $s=0$. In plots (a) and (b) the pixel size is 50 $\mu$m and in plots (c) and (d) it is 100  $\mu$m. For plot (a) and (c) the error on the beam size measurement is 5\%, for plot (b) and (d) it is 10\%. The screens used are assumed to satisfy equation~\ref{eq:cond_theta} and thus the effect of the scattering is neglected in these plots.
\label{fig:errorEmitvsEmit} }
\end{center}
\end{figure*}

Figure~\ref{fig:errorEmitvsEmit} shows the difference between the deducted emittance and the real emittance for several simulation runs.

\begin{figure*}[htbp]
\begin{center}
\begin{tabular}{cc}
(a) Pixels = 50 $\mu$m & (b) Pixels = 100 $\mu$m \\
Noise = 5\% & Noise = 10\% \\
\includegraphics[height=4cm]{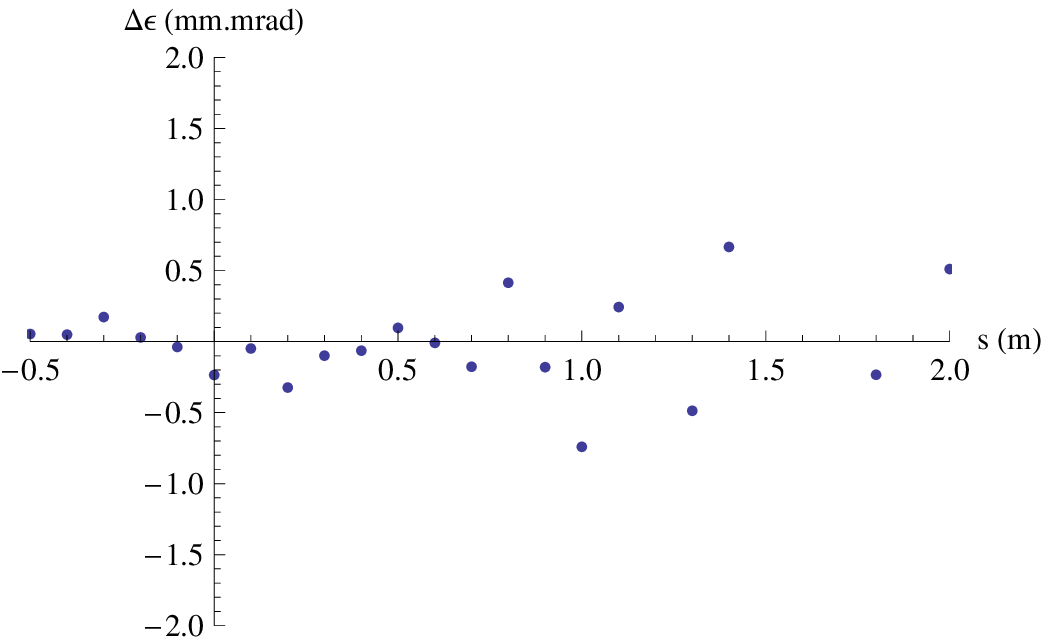} 
& 
\includegraphics[height=4cm]{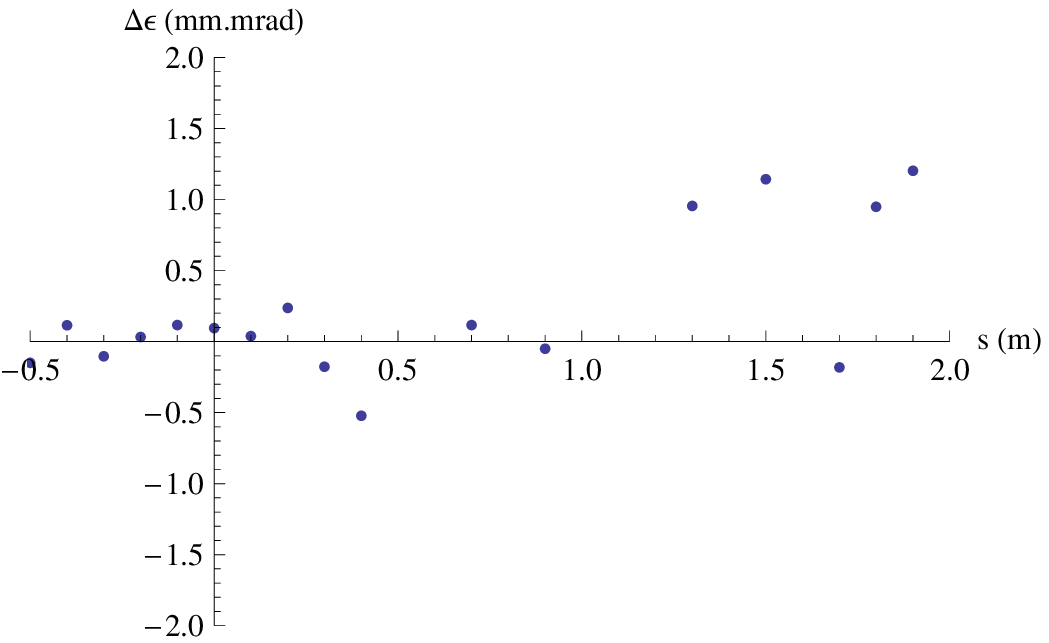} 
\\
(c) & (d) \\
\includegraphics[height=4cm]{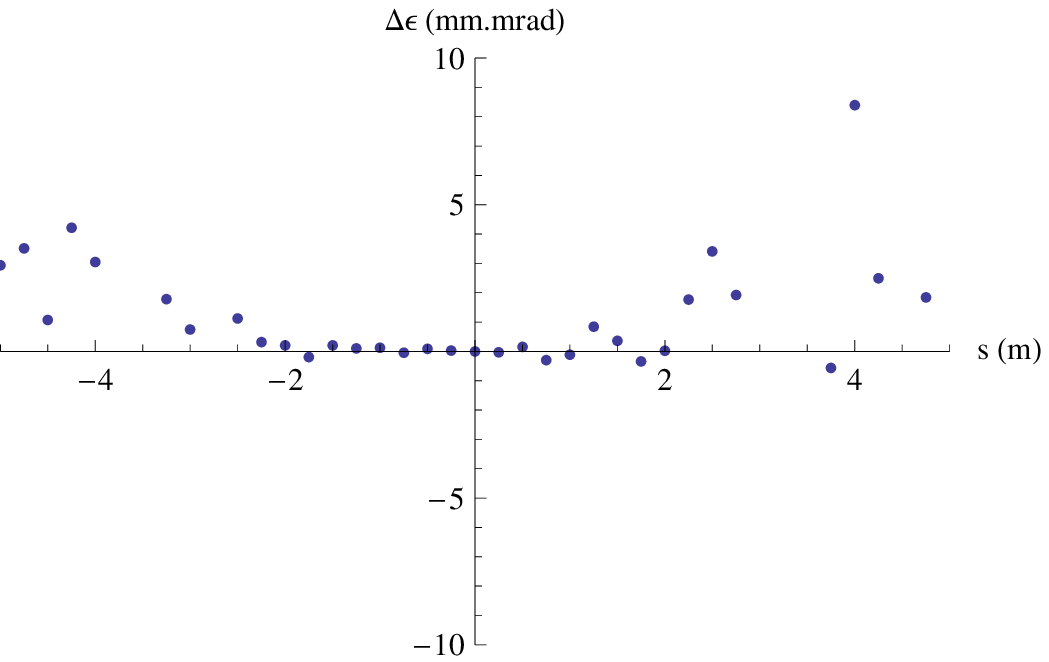} 
& 
\includegraphics[height=4cm]{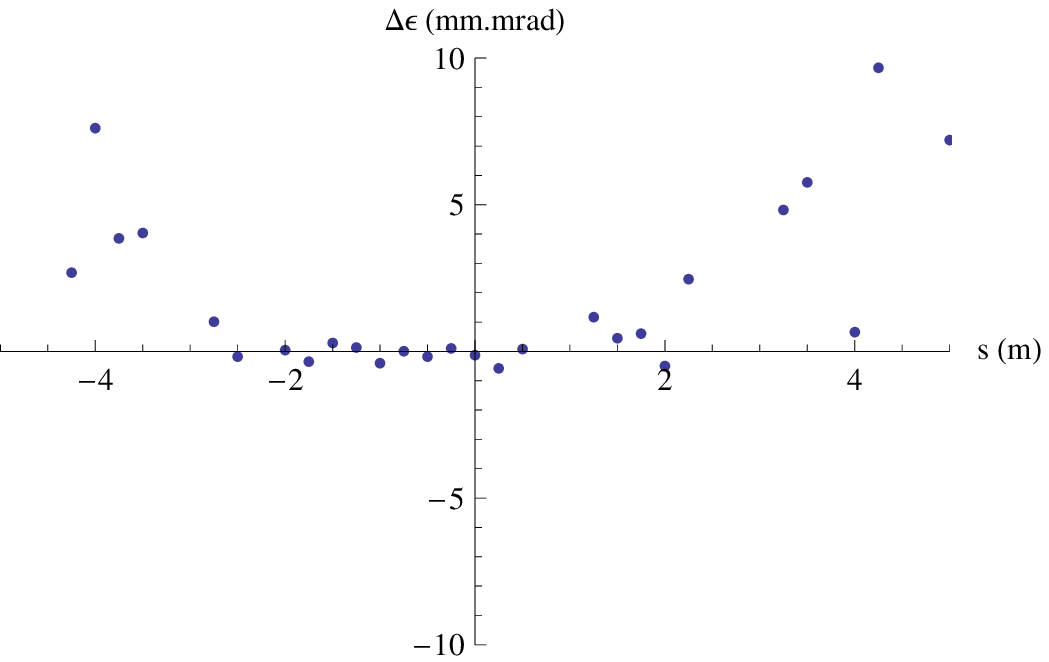} 
\\
\end{tabular}
\caption{Error on the emittance measurement as a function of the position of the beam waist for beams with an emittance of 1~mm.mrad forming a 1~mm waist at the indicated position. In plot (a) and (c) the pixel size is 50 $\mu$m and the noise is 5\%. In plot (b) and (d) the pixel size is 100 $\mu$m and the noise is 10\%. Plot (c) and (d) repeat plat (a) and (b) over a longer range. The screens used are assumed to satisfy equation~\ref{eq:cond_theta} and thus the effect of the scattering is neglected in these plots.
\label{fig:errorEmitvsWaistPos} }
\end{center}
\end{figure*}

Figure~\ref{fig:errorEmitvsWaistPos} shows the difference between the deducted emittance and the real emittance for several simulation runs where the waist position has been varied. These simulations show when the actual waist is far from $s=0$ the performances of the fit are significantly degraded (this is also visible on figure~\ref{fig:emitFits}c).

\subsection{Geant4} 

To better simulate the effect of the scattering in the screens we have used the Geant4 Toolkit~\cite{Agostinelli:2002hh}.

We have used the Geant Low Energy Physics processes~\cite{Apostolakis:1999bp}. These processes are valid from 250eV to 100GeV~\cite{Chauvie:2001fh} and so completely cover the range of energies we are interested in.

In our simulations the electrons were fired from a source on screens placed 0.5m, 1.0m, 1.5m and 2.5m after the source. The initial direction of the electrons was so that they formed a 2mm waist 1m after the source. A 10$\mu m$-thick Mylar window was located 1mm after the electron source.

The position at which the electrons traversed each screen as well as intermediate detectors was recorded. For each simulation more approximately 10~000 electrons were fired.

\subsubsection{Effect of the air pressure}

\begin{figure*}[htbp]
\begin{center}
\hspace*{-2cm}\begin{tabular}{cc}
\includegraphics[height=6cm]{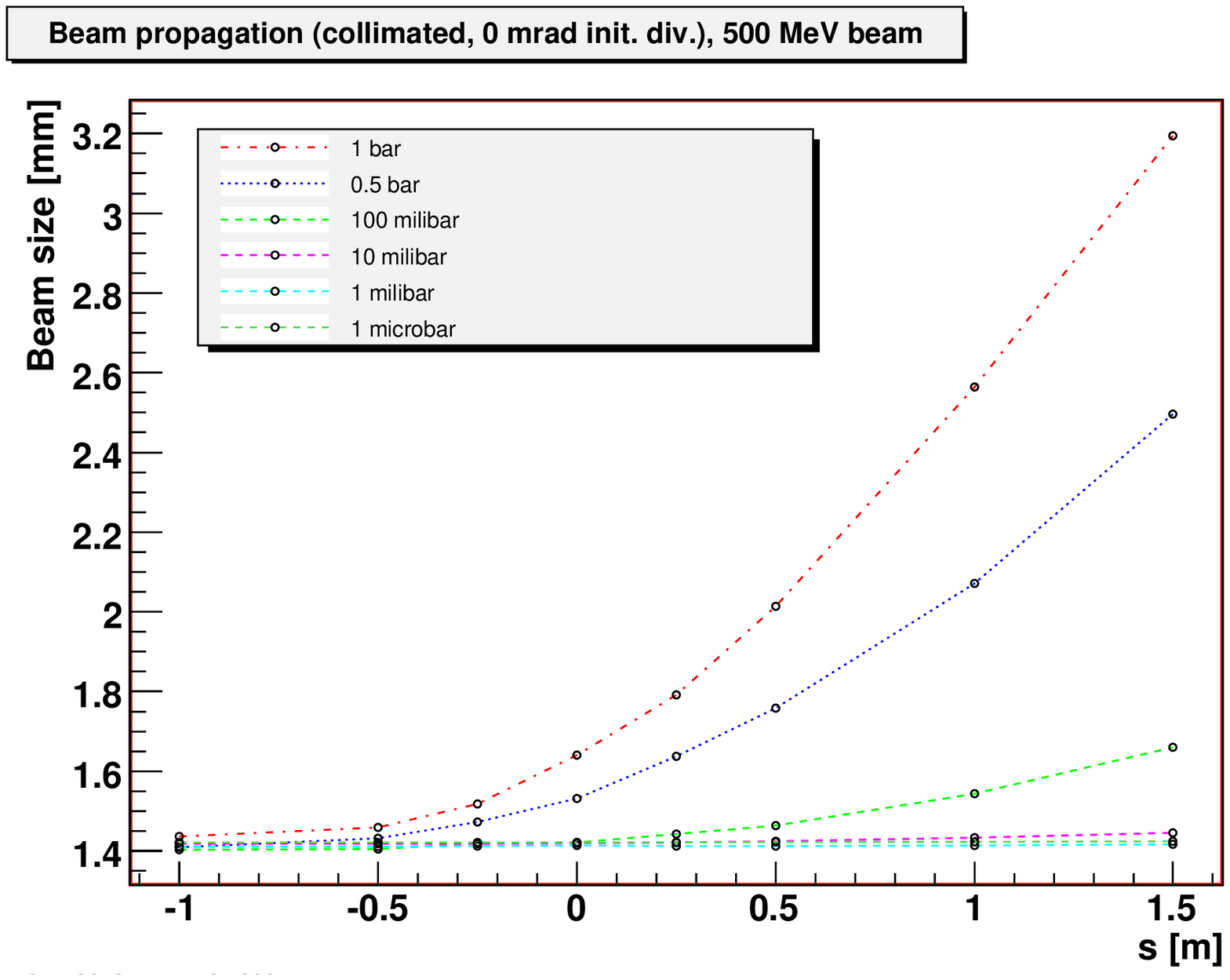}
&
\includegraphics[height=6cm]{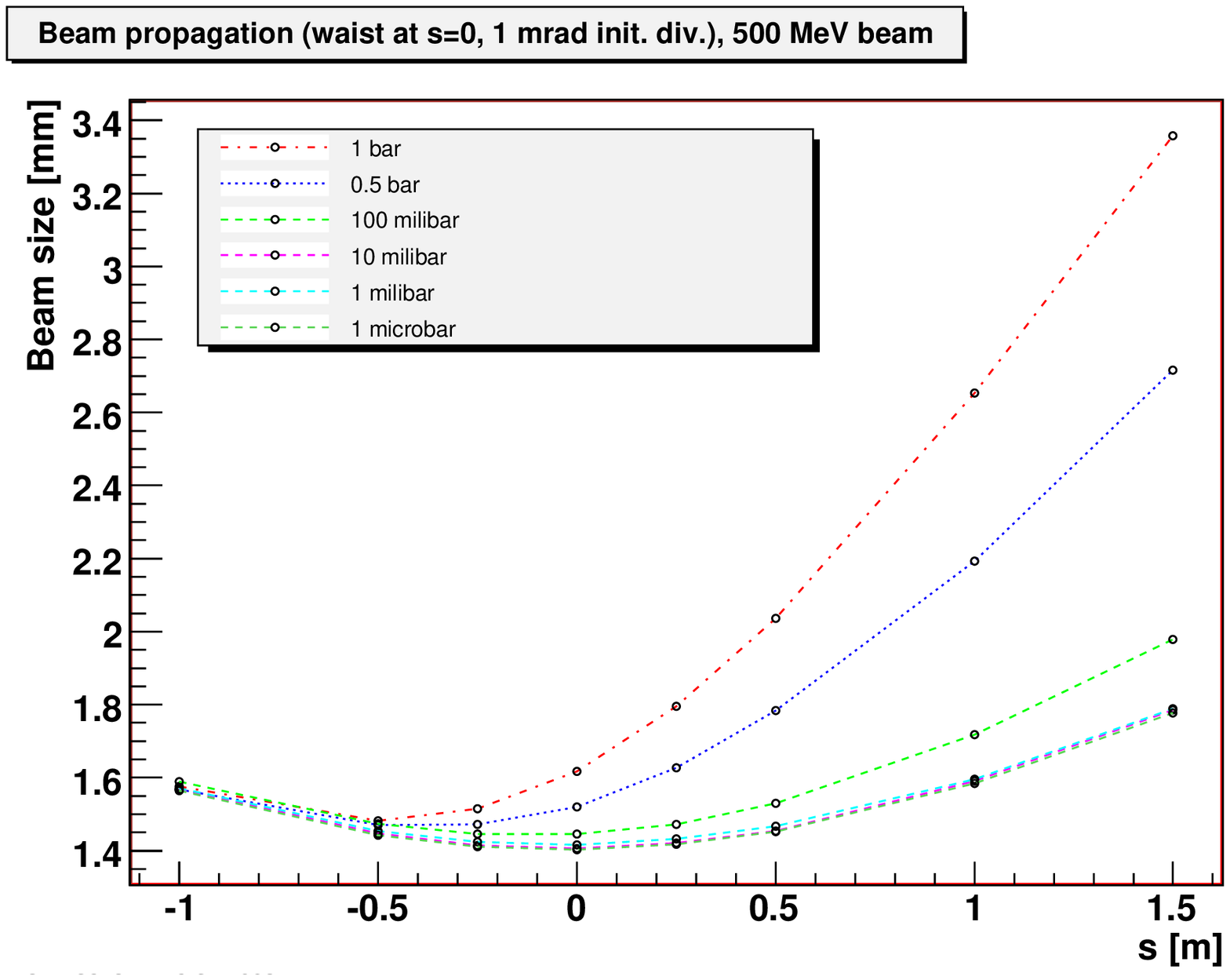}
\\
\end{tabular}
\caption{
Scattering experienced by a 500~MeV beam propagating in various air pressure. On the left plot the beam had no initial divergence whereas the beam on the right  was focused to make a waist at s=0m with a divergence of 1mrad.
\label{fig:scatteringPressure} }
\end{center}
\end{figure*}

Our first study focused on the effect of scattering in the air contained in the vacuum vessel. The propagation of the beam in the absence of screen in an atmosphere filled with air at different pressures was studied. Figure~\ref{fig:scatteringPressure} shows that for pressure higher than 10~mbar the scattering over 2m is significant and thus can not be neglected.

\subsubsection{Emittance measurement}

\begin{figure*}[htbp]
\begin{center}
\hspace*{-2cm}\begin{tabular}{cc}
\includegraphics[height=6cm]{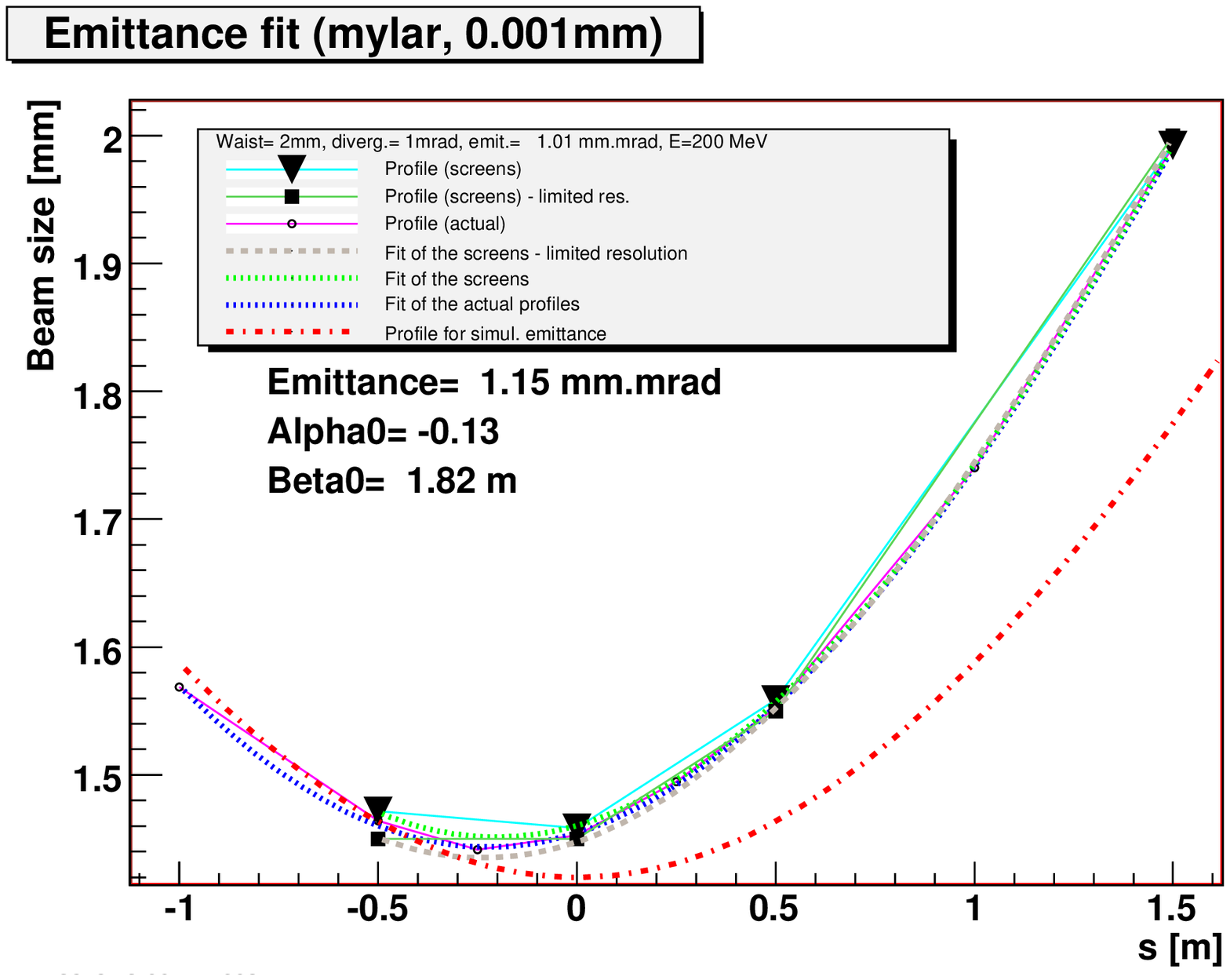} 
& 
\includegraphics[height=6cm]{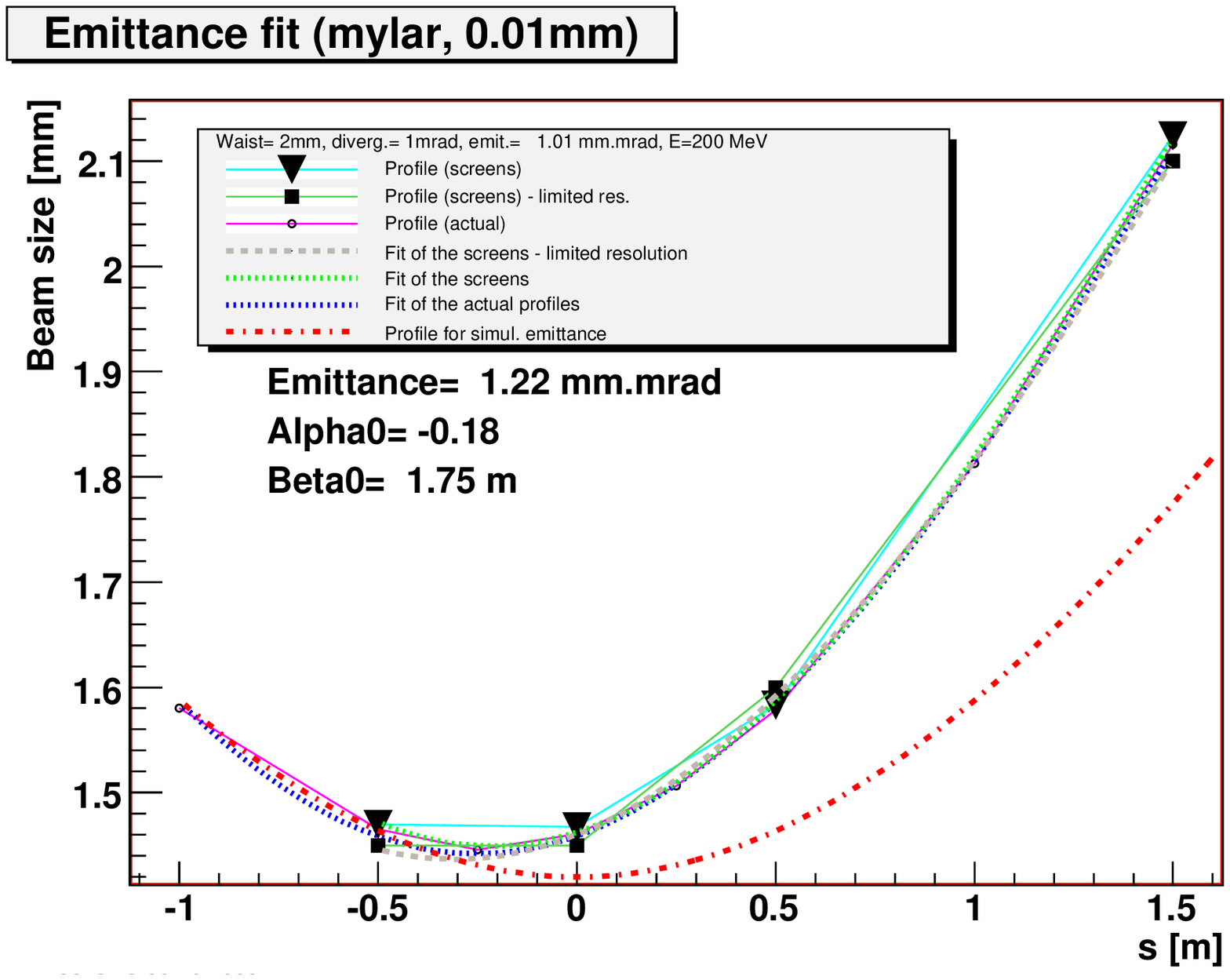} 
\\
\includegraphics[height=6cm]{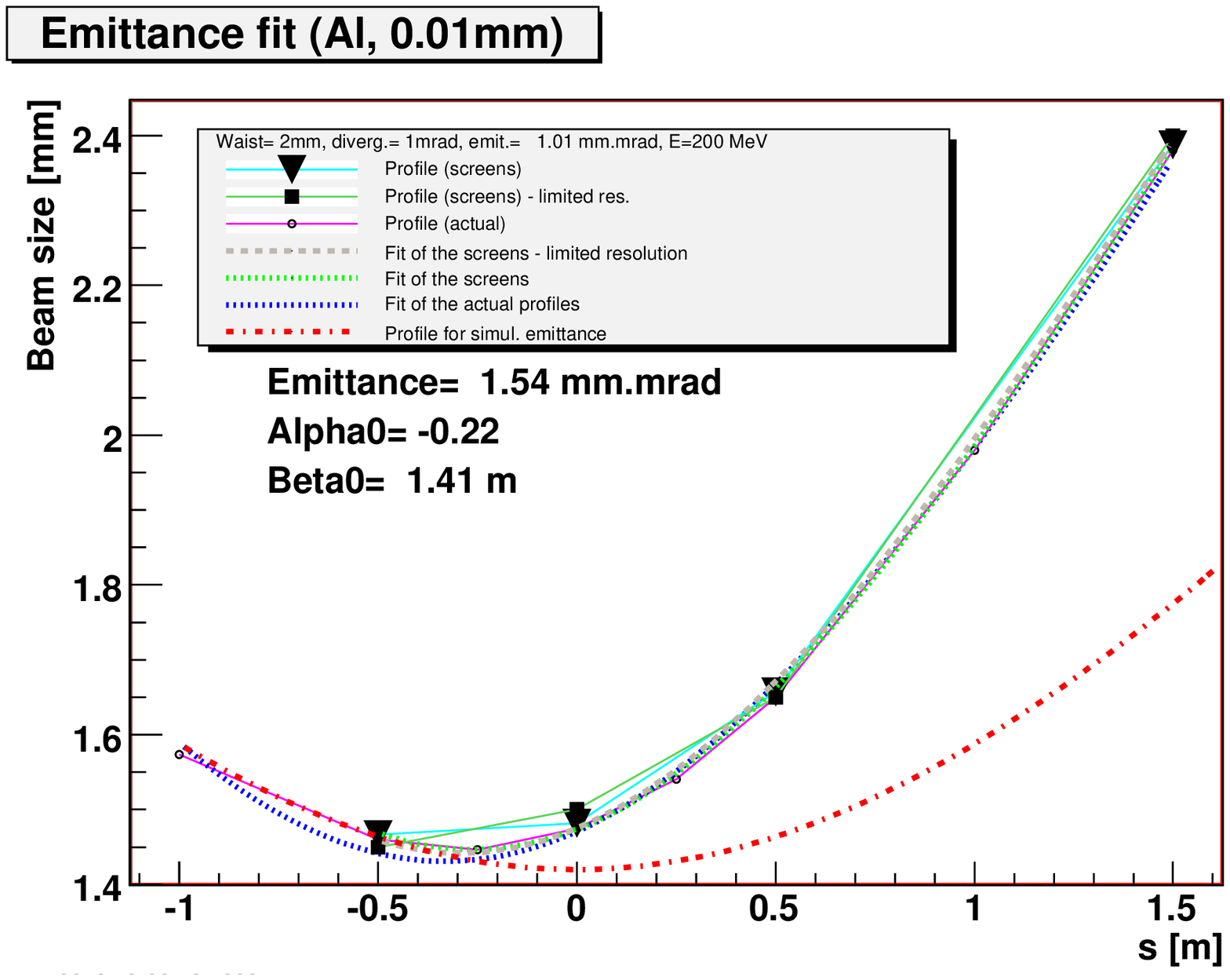} 
& 
\includegraphics[height=6cm]{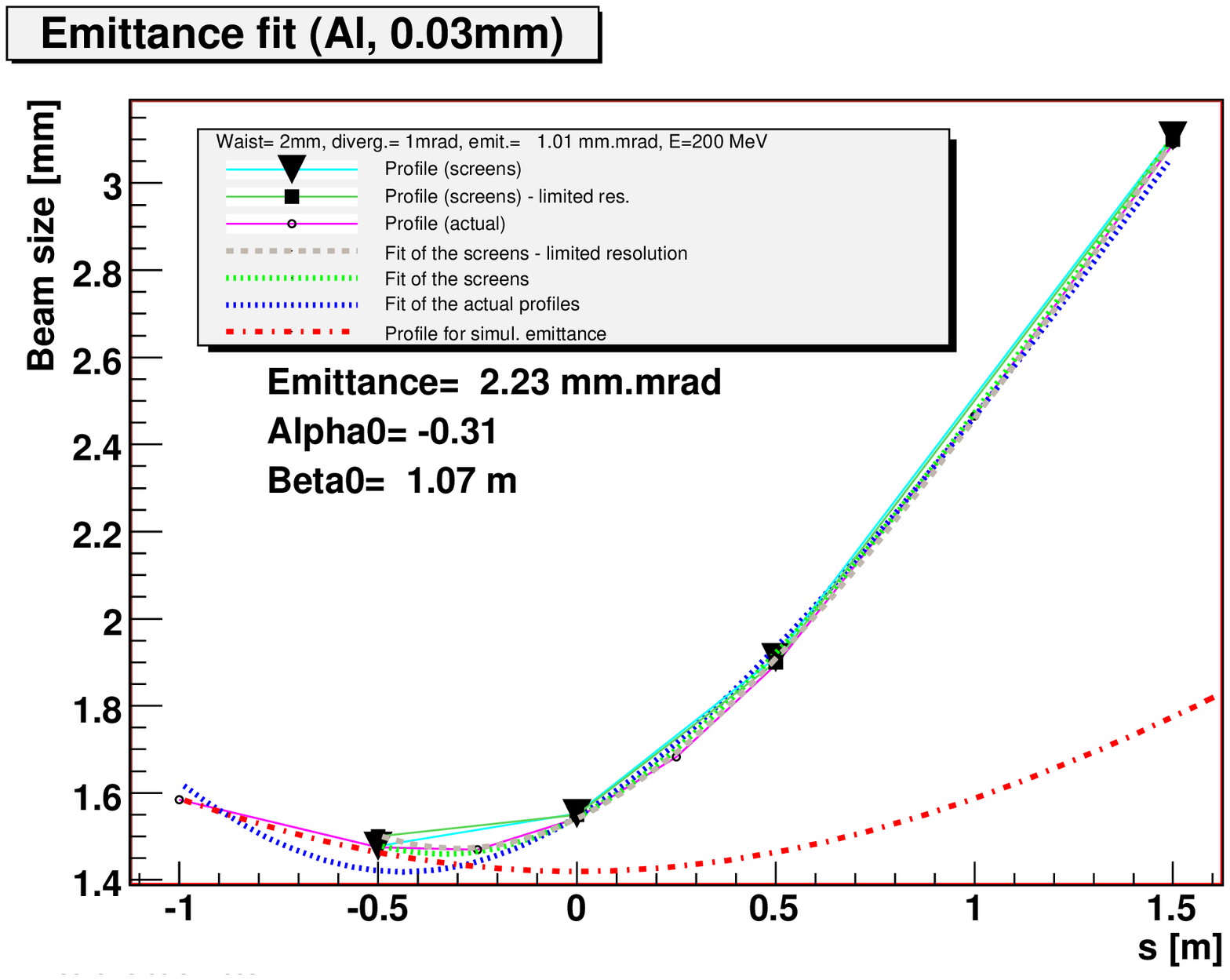}
\\
\end{tabular}
\caption{
Scattering induced in a 200~MeV for various screen material and thickness with an initial emittance of 1mm.mrad. The screens material and thickness is given in the plot title.
The black squares show the beam size observed on each screen, the black triangles show the beam size rounded to the nearest 50 $\mu m$ (to simulate what is observed by a camera with a limited pixel size). The open circles are the position measured at additional intermediate positions.
The dashed brown line shows the emittance fit made using the black triangles. The dashed green line shows the emittance fit made using the black square.  The dashed green line shows the emittance fit made using the open circles.
The dashed red line is the theoretical profile of the beam assuming $\alpha_0=0$, $\beta_0=2.0m$ and $\epsilon=1.01 mm.mrad$. The values given on the plot below the legend for the emittance, $\alpha_0$ and $\beta_0$ are those extracted from the fit on the beam sizes with limited pixel size.
\label{fig:scatteringGeant200MeV} }
\end{center}
\end{figure*}

\begin{figure*}[htbp]
\begin{center}
\hspace*{-2cm}\begin{tabular}{cc}
\includegraphics[height=6cm]{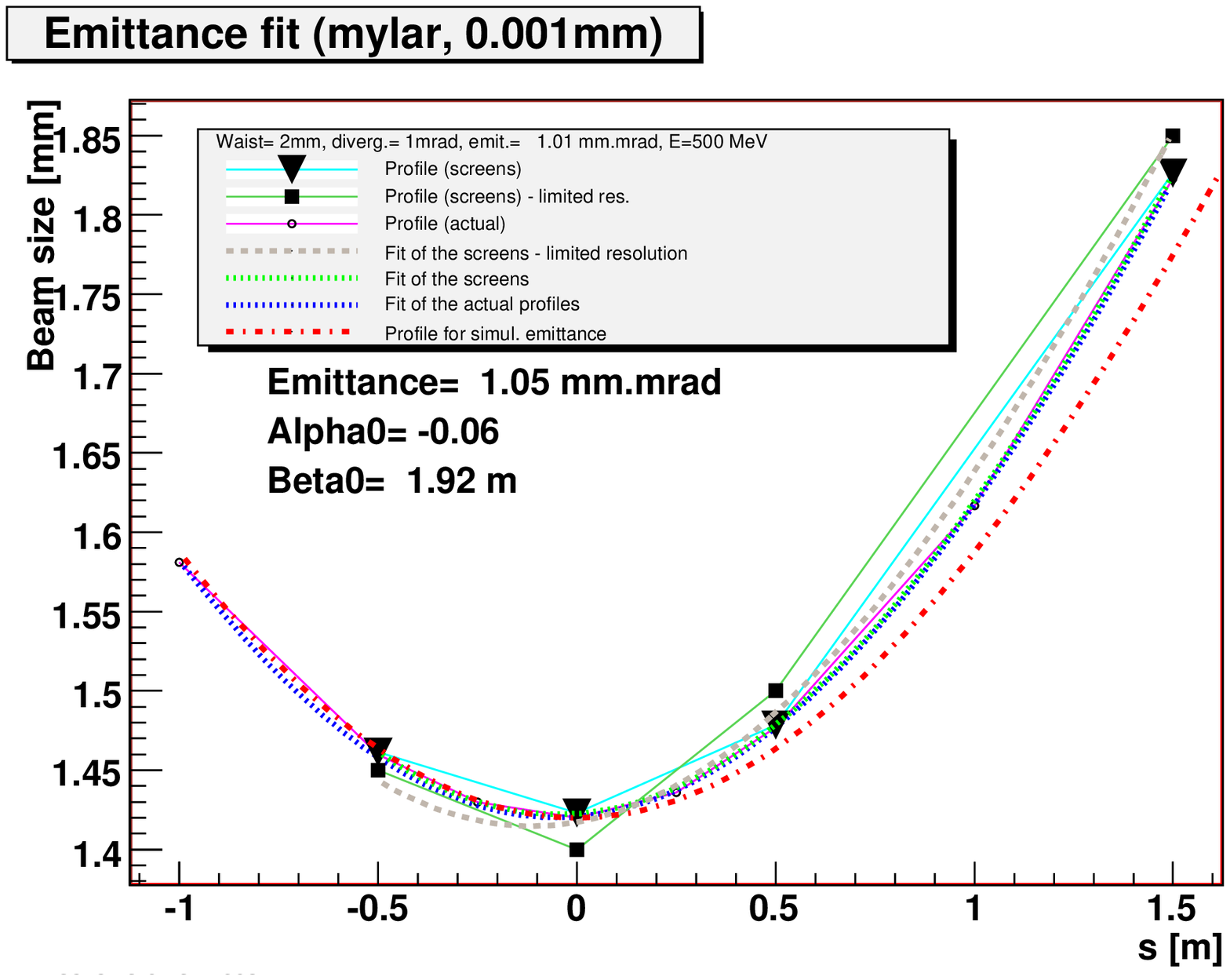} 
& 
\includegraphics[height=6cm]{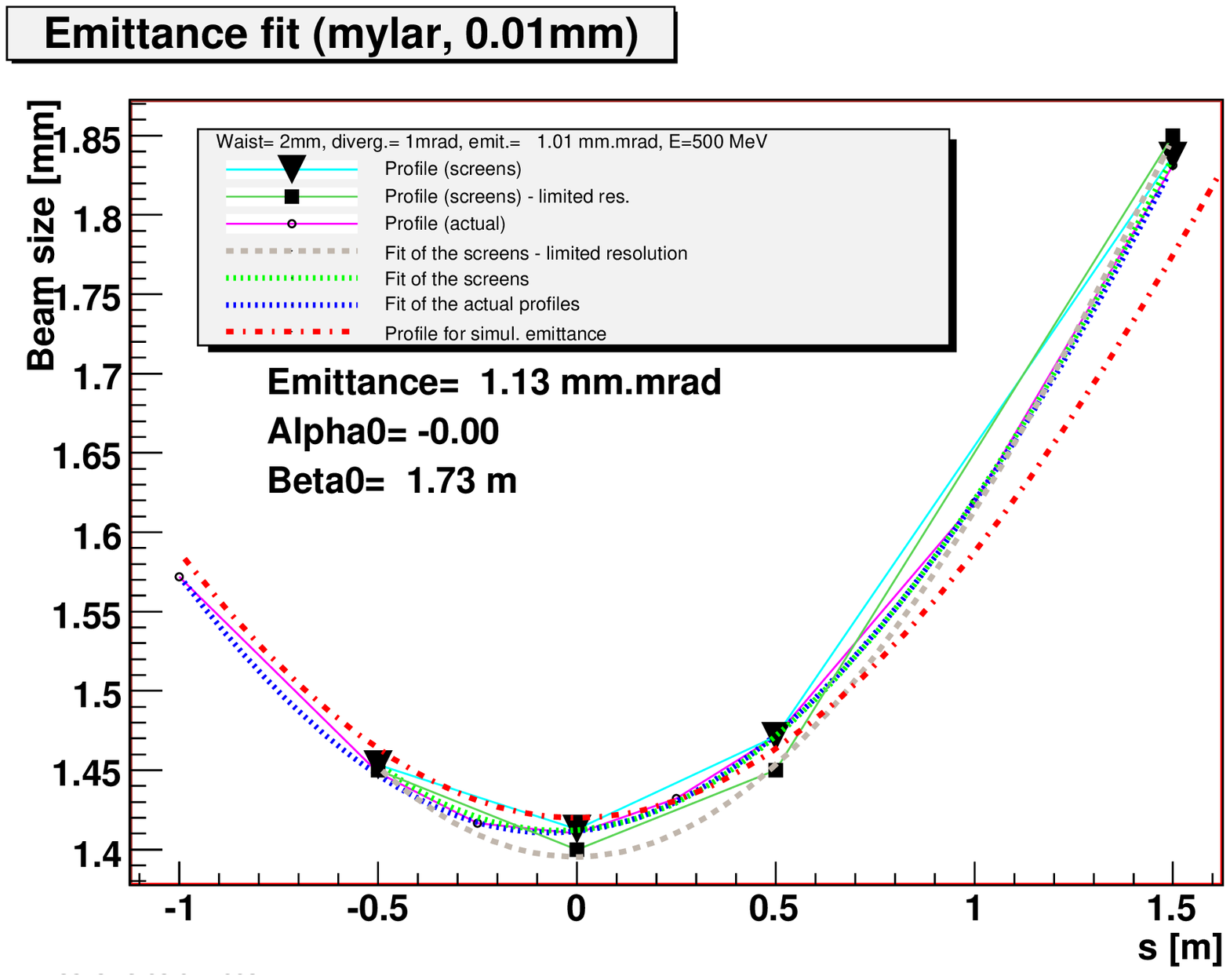} 
\\
\includegraphics[height=6cm]{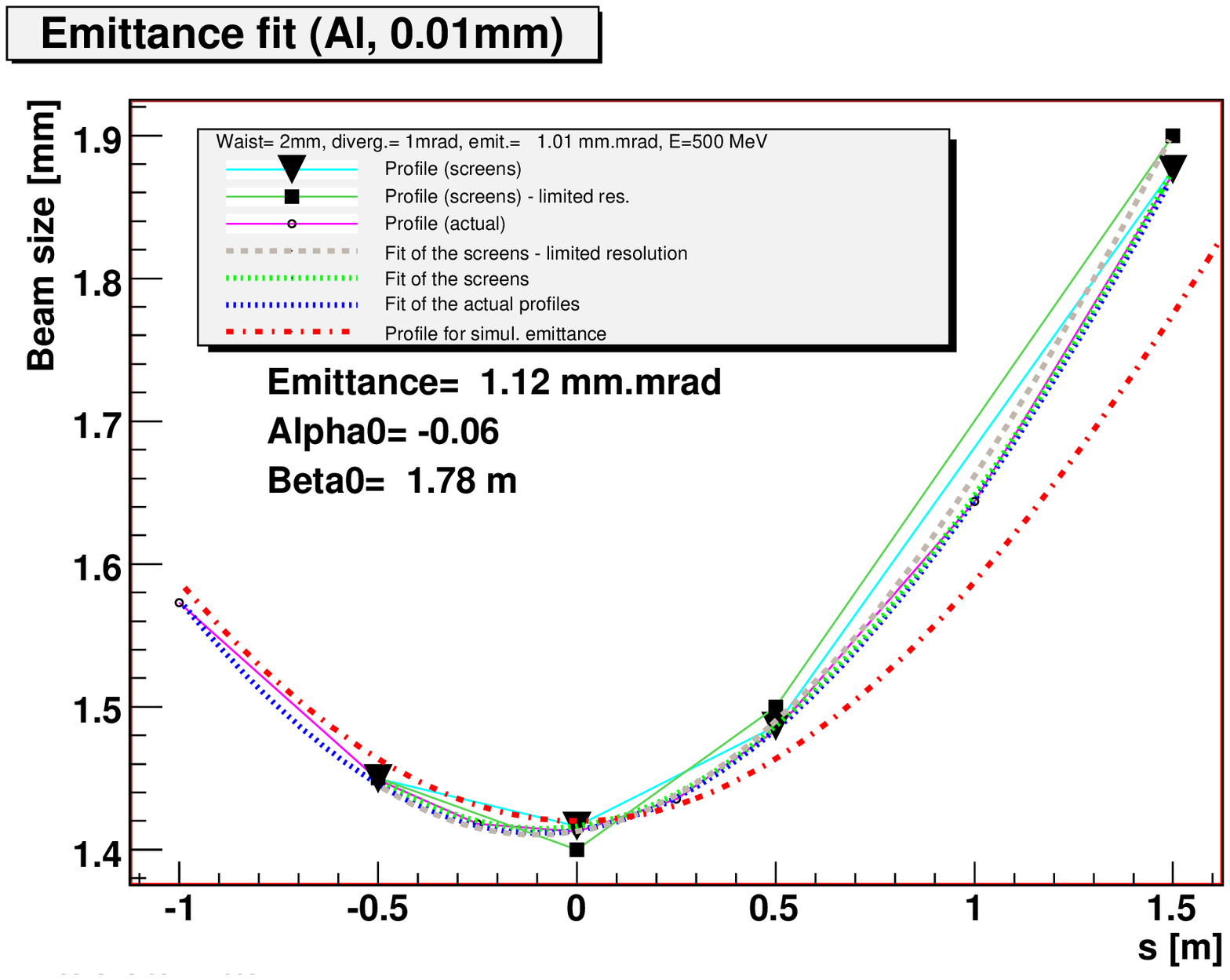}
& 
\includegraphics[height=6cm]{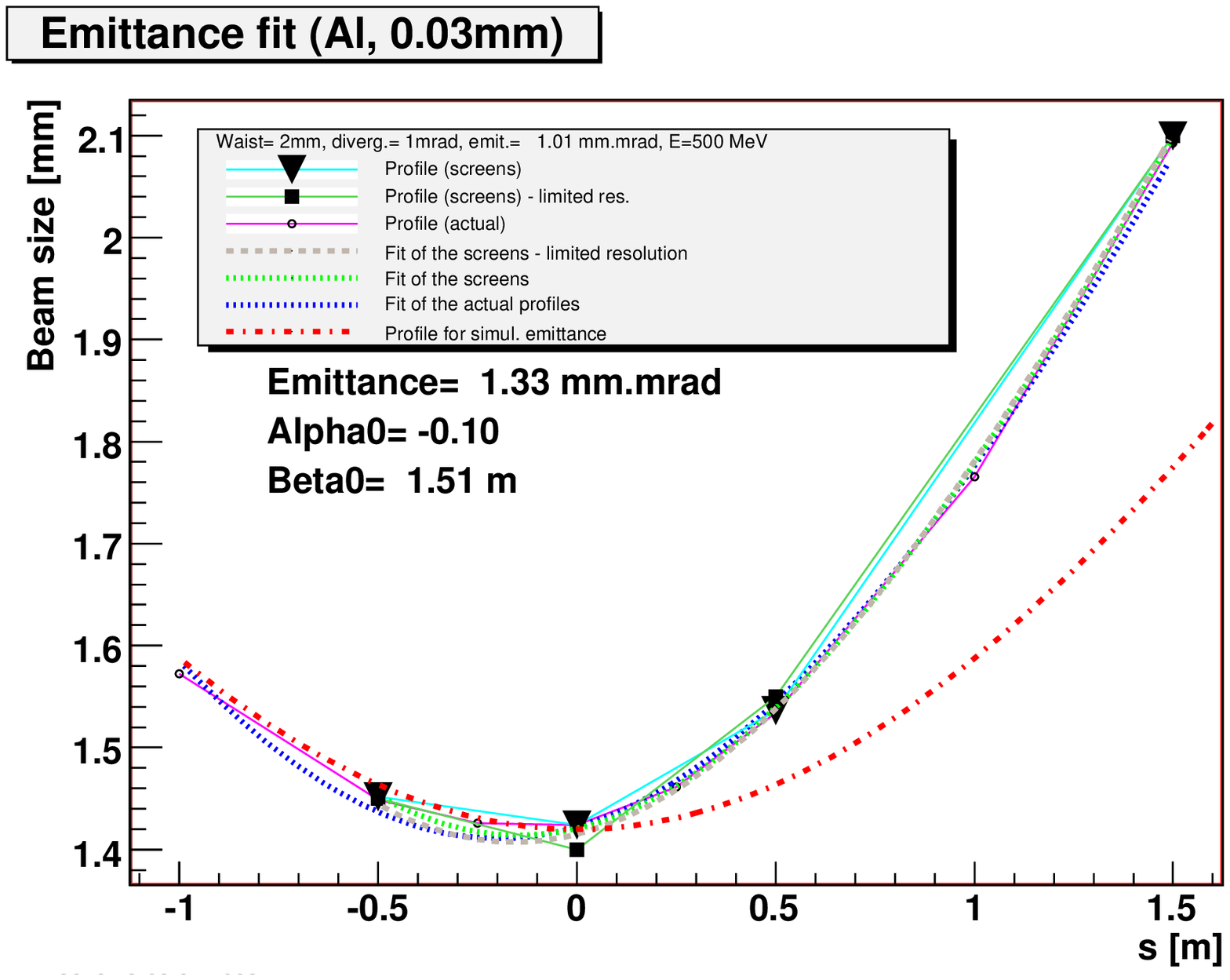} 
\\
\end{tabular}
\caption{
Scattering induced in a 500~MeV for various screen material and thickness with an initial emittance of 1mm.mrad. The screens material and thickness is given in the plot title.
The black squares show the beam size observed on each screen, the black triangles show the beam size rounded to the nearest 50 $\mu m$ (to simulate what is observed by a camera with a limited pixel size). The open circles are the position measured at additional intermediate positions.
The dashed brown line shows the emittance fit made using the black triangles. The dashed green line shows the emittance fit made using the black square.  The dashed green line shows the emittance fit made using the open circles.
The dashed red line is the theoretical profile of the beam assuming $\alpha_0=0$, $\beta_0=2.0m$ and $\epsilon=1.01 mm.mrad$. The values given on the plot below the legend for the emittance, $\alpha_0$ and $\beta_0$ are those extracted from the fit on the beam sizes with limited pixel size.
\label{fig:scatteringGeant500MeV} }
\end{center}
\end{figure*}

To study the effect of the scattering in the screens several simulations were done using different materials and thicknesses for the screens. Some of the simulations for beams of 200~MeV are shown in figure~\ref{fig:scatteringGeant200MeV} and for 500~MeV beam in figure~\ref{fig:scatteringGeant500MeV}. As one can see on the figures the use of thin Mylar screens leads to better results than with Aluminium. Furthermore thin Mylar films are easier to handle.

\begin{figure*}[htbp]
\begin{center}
\hspace*{-2cm}\begin{tabular}{ccc}
(a) -0.5m &
(b) +0.5m &
(c) +1m 
\\
\includegraphics[height=4cm]{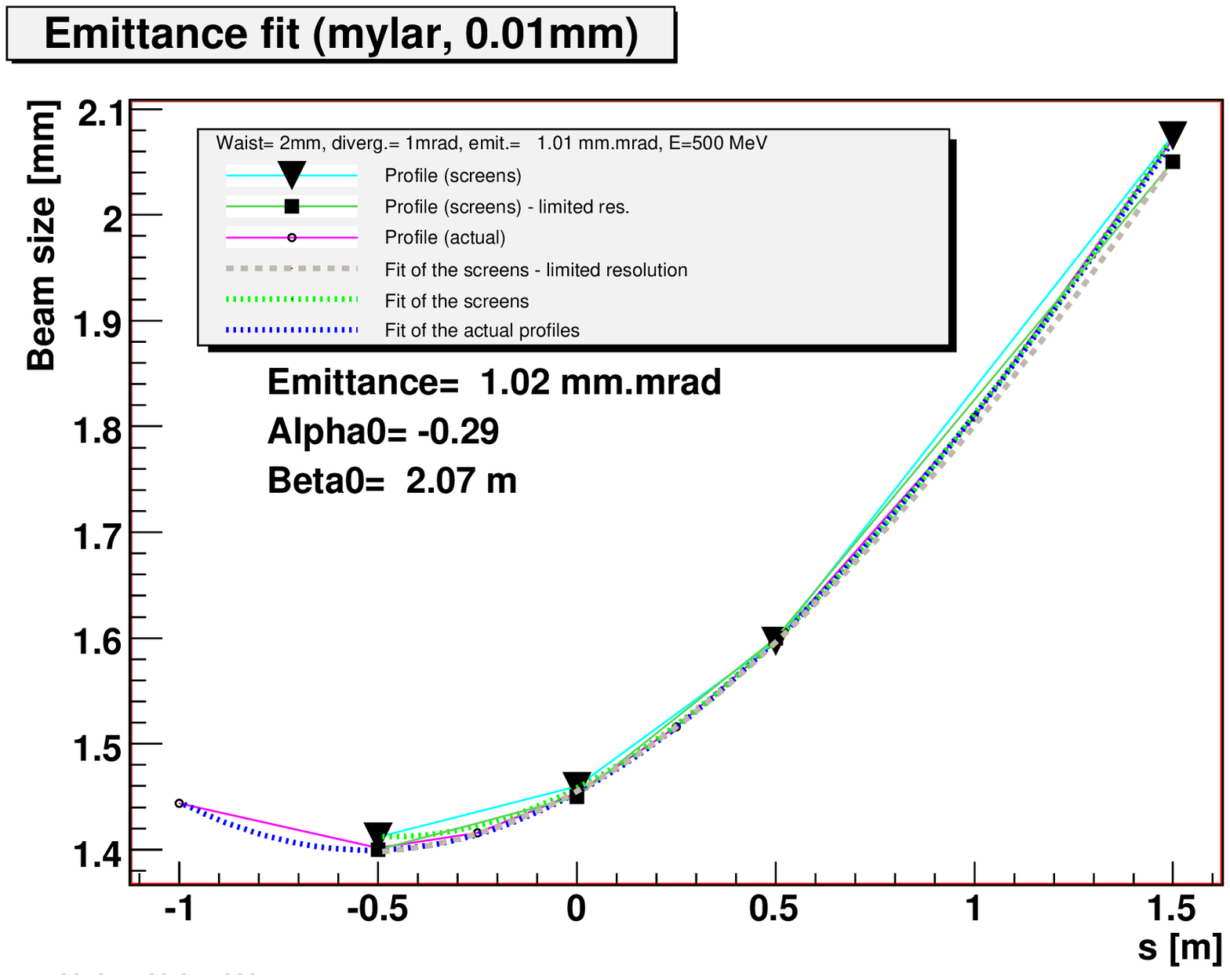} 
& 
\includegraphics[height=4cm]{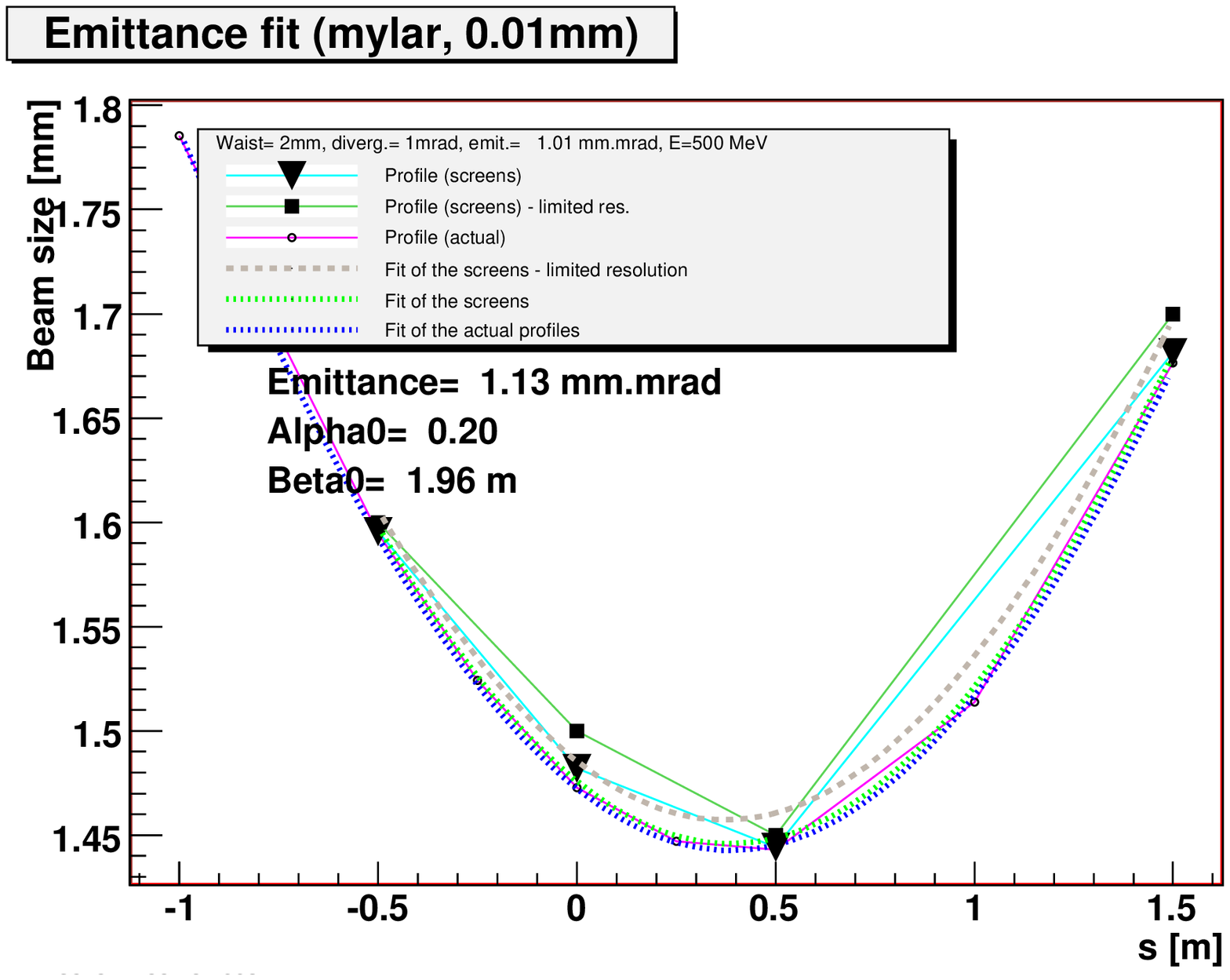} 
& 
\includegraphics[height=4cm]{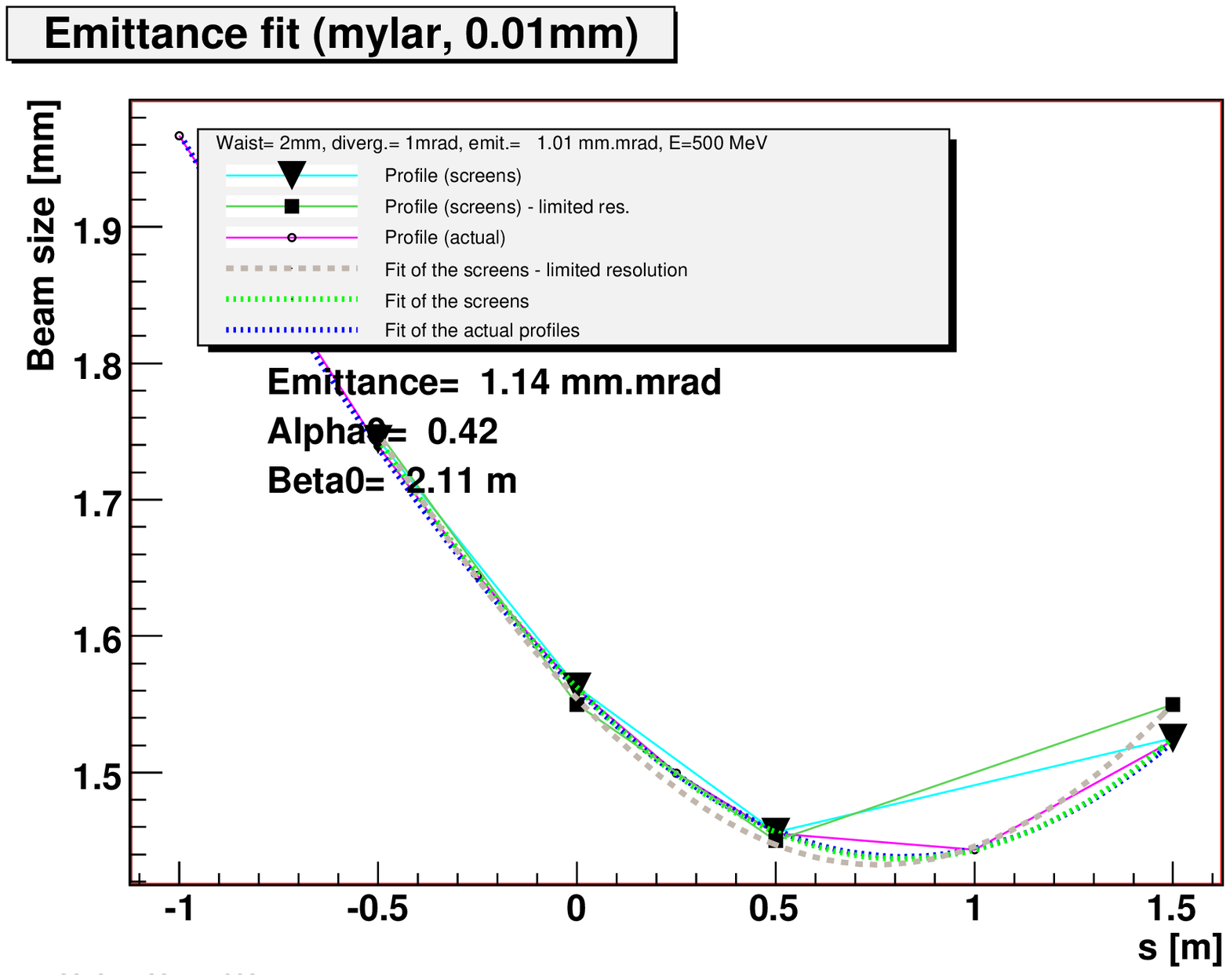} 
\\
\end{tabular}
\caption{
Scattering induced in a 500~MeV for $10 \mu m$-thick Mylar screens for a 500~MeV beam with an initial emittance of 1mm.mrad when the waist is not located on the second screen. The waist position is given above the plot.
The black squares show the beam size observed on each screen, the black triangles show the beam size rounded to the nearest 50 $\mu m$ (to simulate what is observed by a camera with a limited pixel size). The open circles are the position measured at additional intermediate positions.
The dashed brown line shows the emittance fit made using the black triangles. The dashed green line shows the emittance fit made using the black square.  The dashed green line shows the emittance fit made using the open circles.
The values given on the plot below the legend for the emittance, $\alpha_0$ and $\beta_0$ are those extracted from the fit on the beam sizes with limited pixel size.
\label{fig:scatteringGeant500MeVwaistScan} }
\end{center}
\end{figure*}

The previous studies assumed that the beam waist was located on the second screen. Figure~\ref{fig:scatteringGeant500MeVwaistScan} shows that the emittance can still be reconstructed with a reasonable accuracy when the waist is elsewhere (between the screens).

\begin{figure*}[htbp]
\begin{center}
\hspace*{-2cm}\begin{tabular}{ccc}
(a) 200 MeV &
(b) 500 MeV \\
\includegraphics[height=6.5cm]{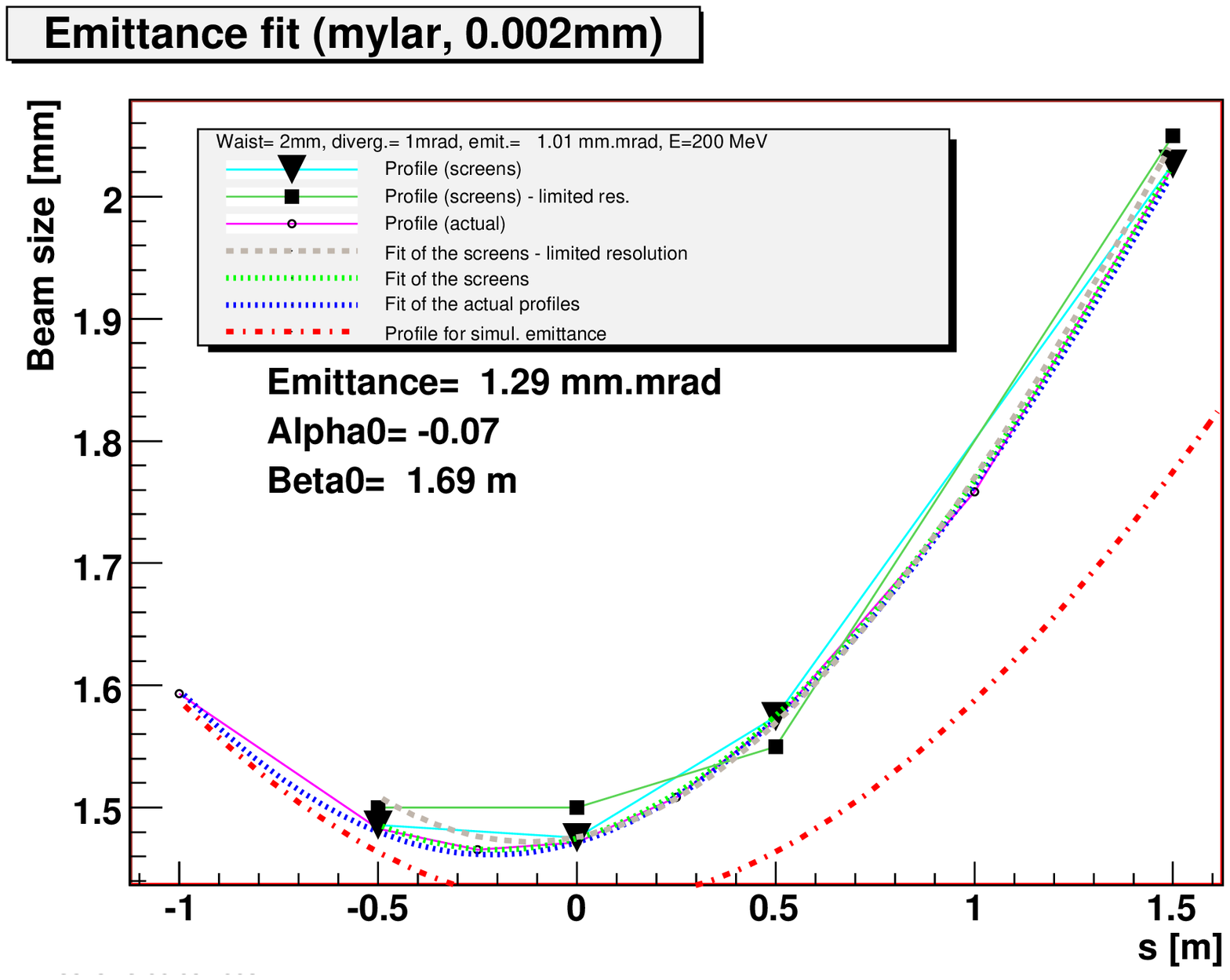} 
& 
\includegraphics[height=6.5cm]{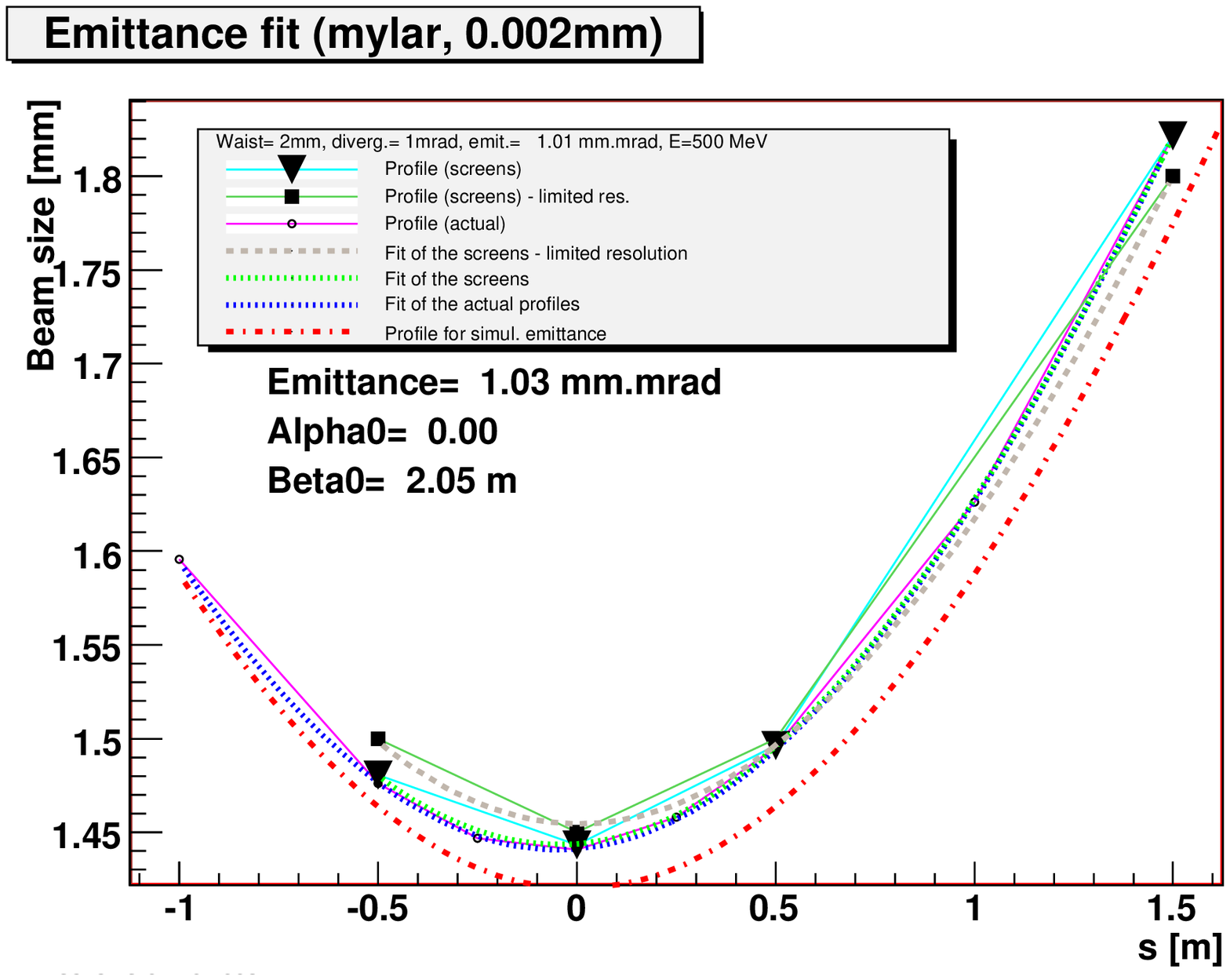} 
\\
\end{tabular}
\caption{
Emittance fit using 2$\mu m$-thick Mylar screens at 200 MeV (a) and 500 MeV (b).
The black squares show the beam size observed on each screen, the black triangles show the beam size rounded to the nearest 50 $\mu m$ (to simulate what is observed by a camera with a limited pixel size). The open circles are the position measured at additional intermediate positions.
The dashed brown line shows the emittance fit made using the black triangles. The dashed green line shows the emittance fit made using the black square.  The dashed green line shows the emittance fit made using the open circles.
The dashed red line is the theoretical profile of the beam assuming $\alpha_0=0$, $\beta_0=2.0m$ and $\epsilon=1.01 mm.mrad$. The values given on the plot below the legend for the emittance, $\alpha_0$ and $\beta_0$ are those extracted from the fit on the beam sizes with limited pixel size.
\label{fig:scattering2umMylar} }
\end{center}
\end{figure*}

Commercially we found some $2 \mu m$-thick Mylar coated with a 10nm layer of Aluminium. Neglecting the scattering in the aluminium, the emittance fit can be seen on figure~\ref{fig:scattering2umMylar}.

%
%
%

\section{Conclusions}

We have studied how 4 screens introduced simultaneously in a beam line could be used to fit the Twiss parameters of an electron beam in a single shot measurement. We have shown that by using $2 \mu m$-thick Mylar screen such fit would lead to a measurement of the beam transverse emittance correct within 15\% for a 500~MeV beam and within 30\% for a 200~MeV beam.

\bibliographystyle{unsrt}
\bibliography{biblio}

\end{document}